\documentclass[aps,pra,twocolumn,groupedaddress]{revtex4-1}
\usepackage{graphicx}
\usepackage{amsmath}
\usepackage{amsfonts}
\usepackage{color}
\usepackage[utf8]{inputenc}
\usepackage[OT4]{fontenc}

 \def\urlprefix{}
 \def\url#1{}
\begin{document}
\renewcommand*{\figurename}{FIG.}
\newtheorem{thm}{Theorem}
\newtheorem{prop}{Proposition}
\newenvironment{proof}[1][\textit{Proof}]{\begin{trivlist}
\item[\hskip \labelsep {#1}]}{\end{trivlist}}

\def\be{\begin{equation}}
\def\ee{\end{equation}}
\def\bea{\begin{eqnarray}}
\def\eea{\end{eqnarray}}
\def\bi{\begin{itemize}}
\def\ei{\end{itemize}}
\def\bin{\begin{enumerate}}
\def\ein{\end{enumerate}}
\def\la{\langle}
\def\ra{\rangle}
\newcommand{\mv}[1]{\left\la#1\right\ra}
\newcommand{\vect}[1]{\mathbf{#1}}

\title{{Variational Bose-Hubbard model revisited}}

\author{Jan Major$^{1}$, Mateusz \L\k{a}cki$^{1}$, Jakub Zakrzewski$^{1,2}$}

\affiliation{ $^1$Institute of Physics, Jagiellonian University, ul.\
Reymonta 4, 30-059 Krak\'ow, Poland
 \\$^2$Mark Kac Complex Systems Research Center, Jagiellonian University, ul.\ 
Reymonta 4, 30-059 Krak\'ow, Poland}


\begin{abstract}
For strongly interacting bosons in optical lattices the standard description using
Bose-Hubbard model becomes questionable. The role of excited bands becomes important.
In such a~situation we   compare results of simulations  using multiband Bose-Hubbard model with a~recent proposition based on a~time dependent variational approach. It is shown that the latter, in its original formulation, uses too small variational space leading often to spurious effects. Possible expansion of variational approach  is discussed. 
\end{abstract}

\pacs{67.85.Hj,03.75.Kk,03.75.Gg,03.65.Ud}

\maketitle

\section{Introduction}

Ultracold bosonic atoms in an optical lattice potential have been a~very active field of both experimental and theoretical research. They enable to prepare and study a~broad spectrum of complex quantum systems in well-controllable experiments. Of particular importance is experimental research, using ultra cold quantum gases, of systems which mimic various condensed matter Hamiltonians. For a~complete review please see \cite{Lewenstein2007,Lewenstein12}. The key stimulus for these activities is existence of the mapping, from a~continuous model \cite{Gersch1963}, describing a~gas of ultra cold atoms in a~optical lattice potential \cite{Jaksch1998} to a~discrete Bose-Hubbard model.

By choosing a~proper setup of lasers forming an optical lattice, various lattice geometry and dimensionality may be realized \cite{Jessen1996,Becker2010}. In particular one may obtain a~one-dimensional lattice \cite{Stoferle2004}, which is then mapped to a~one-dimensional Bose-Hubbard model. One dimension makes interactions and correlations relatively strong \cite{Kuhner1998,Giamarchi2004}, driving the system far from the mean field regime\cite{Fisher1989}. The mapping is performed by expanding the field operator in localized modes with the help of  Wannier functions \cite{Kohn1959,Kivelson1982}. While often restriction to the lowest Bloch band, leading to a~standard Bose-Hubbard model \cite{Jaksch1998} is sufficient, for stronger interactions
higher bands become significant \cite{Oosten2003,Diener2006,Koetsier2006,Mering2011,Bissbort2012,Luehmann2012,Lacki2013a}. 
Including them enlarges the dimension of the local Hilbert space describing configuration of particles within one lattice site. This makes the numerical analysis  computer time demanding.

Recently, a~new proposal for an approximation, designed to address that particular issue, has been made \cite{Sakmann2011}. Authors use a~time-dependent variational principle to optimize a~single one-particle Wannier function per site. Its shape is altered  by interactions  with the other particles  also during the evolution, while in a~standard approach \cite{Jaksch2002} Wannier functions depend solely on the instantaneous strength of the optical lattice potential,i.e. on a~single-particle physics. In the variational approach, a~dynamic change of  Wannier functions may be a~substantial improvement by allowing them to be chosen optimally. The question remains if and under what assumptions this choice (limited nevertheless to the Hilbert space spanned by the variation) is good enough for realistic problems. This is the problem we want to address in this paper.

We discuss the multi-band Bose Hubbard model reduced to one-dimension in Section II 
while Section III brings necessary information concerning the time-dependent variational approach. Comparison of both approaches is given in Section IV both on the ground state and different time-dependent dynamical problems. We restrict ourselves to small model systems that nevertheless allow us to compare both methods. A simple generalization of the vairational approach and its possible advantages is discussed in Section V with the subsequent sections presenting our conclusions.

\section{The Multiband Bose-Hubbard Model}

Ultracold interacting gas of bosons in the optical lattice potential is described by a~second quantized Hamiltonian: 
\bea
 {\hat{H}}&= &\int \textrm{d}^3r \Psi^\dagger(\vec r)\hat{h}(\vec{r})\Psi(\vec r)+  \nonumber\\
&+&\frac{1}{2}\int\textrm{d}^3r \textrm{d}^3r'  \Psi^\dagger(\vec r) \Psi^\dagger(\vec r')V(\vec r, \vec r') \Psi(\vec r) \Psi(\vec r'), 
\label{eq:Hx}
\eea
where $\hat{h}=-\frac{\hbar^2}{2 m } \nabla^2+V_{\mathrm{lat}}(\vec r)$ is a one-particle Hamiltonian and 
\be
V(\vec r,\vec r')=\frac{4\pi \hbar^2 a}{m}\delta^{(3)}(\vec r-\vec r')=g\delta^{(3)}(\vec r-\vec r')
\label{eqn:regpot}
\ee
 is a~contact pseudopotential modelling s-wave scattering interaction with $a$ being the scattering length. Formally, to avoid problems 
 with  hermiticity of the above Hamiltonian \cite{Exner2005} instead of Dirac-delta interaction, one should use a~pseudopotential of the form
 \be 
V(\vec r,\vec r')=g\delta(\vec r-\vec r')\frac{\partial}{\partial |\vec r-\vec r'| }|\vec r-\vec r'|.
\label{eqn:regpot2}
\ee
 However in the multiband expansion, one typically uses a~basis spanned by smooth Wannier functions truncated to first few Bloch bands (for details see the next section). In that case the potential (\ref{eqn:regpot2}) is equivalent to the simplified Dirac delta potential (\ref{eqn:regpot}).

In the following we consider a~quasi one-dimensional geometry assuming 
 $V_{lat}(\vec r) = s \sin^2(k x)+\frac{1}{2}m\Omega^2(y^2+z^2),$ where $\Omega$ is a~frequency of a~tight transverse harmonic trapping potential. In these transverse directions we assume that the ground state mode $\phi_0$ is occupied only. For a~given lattice depth $s$, the field operator 
is expanded as  
\be
 \Psi(\vec r)=\sum_{i,\alpha} a_i^\alpha W^\alpha_i(\vec r),
\label{eq:Psi}
\ee
with
\be
W^\alpha_i(\vec r)= w^\alpha_i(x)\phi_0(y)\phi_0(z),
\ee
where $w^\alpha_i(x)$ is the standard one-dimensional Wannier function of $\alpha$ band \cite{Kohn1959} localized at site $i$. Performing integrations in Eq.~(\ref{eq:Hx}) the
multiband  model is obtained:

\begin{eqnarray}
&&\hat{H} = -\sum\limits_{i\ne j,\alpha}J_{i-j}^{\alpha} (\hat{b}_i^{\alpha\dagger} \hat{b}_{j}^\alpha  + h.c.)+ \sum_{i,\alpha} E_i^\alpha \hat{n}_i^\alpha+ \nonumber\\
&&\frac{1}{2}\!\!\sum\limits_{\alpha,\beta,\gamma,\delta}\sum\limits_{ijkl} U_{ijkl}^{\alpha\beta\gamma\delta} \hat{b}^{\alpha\dagger}_i \hat{b}^{\beta\dagger}_j \hat{b}^{\gamma}_k \hat{b}^{\delta}_l.
\label{eqn:ebh}
\end{eqnarray}
The tunnelling from site $j$ to $i$ (along $x$ direction) in the $\alpha$ band is 
\be J_{i-j}^\alpha=\int w^\alpha_{i}(x)\left[ -\frac{\hbar^2}{2m}\frac{d^2}{d x^2}+  s \sin^2(k x)\right]w^\alpha_{j}(x) \textrm{d}x,
\ee 
with mean energies at sites in different bands $E_i^{\alpha}=J_{0}$ being independent of site. Often in experiments additional slowly varying harmonic trap potential is present which may be taken into account in $E_i^{\alpha}$'s. For the purpose of the present work
such terms are not relevant and are dropped for simplicity. 
The interaction integrals read 
\be
U_{ijkl}^{\alpha\beta\gamma\delta} = g\int \textrm{d}x  w_i^\alpha(x)w_j^\beta(x)w_k^\gamma(x)w_l^\delta(x),
\ee
with 
\be g= \frac{4\pi \hbar^2 a}{m}\int \textrm{d}y\textrm{d}z |\phi_0(y)|^4|\phi_0(z)|^4
\ee
being a~modified contact interaction strength due to reduction of the problem to one dimension. In terms of the transverse trap frequency it reads $g=2 \hbar a\Omega$.

For sufficiently deep lattices (with depth $s$ of a~few energy recoils $E_R=\hbar^2k^2/2m$) one may make a~standard approximation neglecting long range tunnelings  $J_{i-j}, $for $|i-j|\geq 2$ and keeping nearest neighbor tunnelings $J_1$ only (later we drop the subscript and denote this tunnelling simply as $J$ following the standard convention). 
Similarly often only the on-site interactions terms $U_{ijkl}^{\alpha\beta\gamma\delta}$ for $(i,j,k,l)=(i,i,i,i)$ are taken into account since other integrals are significantly smaller. Recently, however, it has been stressed \cite{Dutta2011,Mering2011,Luehmann2012,Lacki2013a} that contributions $U_{ijkl}^{\alpha\beta\gamma\delta}$ for $(i,j,k,l)=(i,i,i,j)$ (up to a~permutation), may not be easily dismissed.
They have a~character of a~density-dependent tunnelling and they may compete with standard tunnelings (especially for deep lattices, strong interactions, or large density) leading to significant, measurable effects.

While we could take these terms into account, we choose to neglect them in the following to concentrate on the comparison between multiband and variational approach on a~standard Bose-Hubbard system without density dependent tunnelings, as introduced in \cite{Sakmann2011}.

With these assumptions  the  multiband Bose-Hubbard (MBH) Hamiltonian reads:
\begin{align}
&\hat{H}_{\mathrm{MBH}} = \sum_{k=1}^L\left(-\sum\limits_{\alpha=1}^{\cal N}J^{\alpha} (\hat{b}_k^{\alpha\dagger} \hat{b}_{{k+1}}^\alpha  + h.c.)+ \sum_{\alpha=1}^{\cal N} E_k^\alpha \hat{n}_k^\alpha+ \right.\nonumber\\
&\left.\frac{1}{2}\!\!\sum\limits_{\alpha,\beta,\gamma,\delta}^{\cal N} U^{\alpha\beta\gamma\delta} {\hat{b}_{{k}}^\alpha}{}^\dagger {\hat{b}_{{k}}^\beta}{}^\dagger \hat{b}_{{k}}^\gamma \hat{b}_{{k}}^\delta\right)
\label{eqn:ebh2},
\end{align}
where ${\cal N}$ is a~number of bands taken and we have dropped the subscripts on interaction constants as they become, within
the assumed model, independent on sites. The above  Hamiltonian is used in the simulations  in the subsequent sections. It is also a~basis for formulation of the variational system of equations of motion described in the next section. In this work we restrict ourselves to the analysis of small systems consisting of a~few sites.

\section{Time Dependent Bose-Hubbard Model from variational principle }
\label{sec:TDV}
While forming a~single band Bose-Hubbard (BH) model, a~special case of the MBH, one neglects the contribution from higher Bloch bands. For strong interparticle interactions, this may significantly alter the results. 

There have been attempts at restricting the Hamiltonian (\ref{eqn:ebh2}) to a~relevant Hilbert subspace \cite{Johnson2009,Dutta2011,Mering2011,Luehmann2012,Lacki2013a} by renormalizing single band BH model's parameters to density-dependent values including effectively influence of the higher bands. This approach is suitable only for low energy physics, when excited bands are not populated.

Another interesting variational approach to simulate multiband effects has been proposed originally in \cite{Sakmann2011}. We review its formulation below for self-containment of the paper.  This variational single band model assumes that particles do not populate single particle modes defined by the ordinary Wannier functions, but time-dependent modes  formed by linear combinations of  Wannier functions with appropriate time-dependent coefficients $d_k^\alpha(t)$. For one dimensional system this gives

\begin{align}
w_k(x,t)=\sum^{{\cal N}_{\mathrm{V}}}_{\alpha=1}d^{\alpha}_k(t)w^{\alpha}_k(x),
\label{eqn:wannierSak}
\end{align}
with  $w_k^\alpha(x)$ being the standard (time independent) Wannier functions used also in the previous section. The coefficients are allowed to vary in time and are chosen variationally by the Time-Dependent Variational (TDV)  principle \cite{Dirac1930,Frenkel1934,McLachlan1964,Broeckhove1988}.

The novel idea in this approach is that the dynamics  of Wannier functions $w_k(x,t),$ is set by the variational principle and  not simply determined by e.g. the time dependence of optical lattice potential depth. By construction they are mutually orthogonal and may be assumed to form the orthonormal set: $\langle w_i(t)|w_j(t)\rangle=\delta_{ij}$. Many boson wave function is defined as:
\begin{align} 
|\Psi(t)\rangle=\sum_{\vec{n}}C_{\vec{n}}(t)|\vec{n};t\rangle,
\label{eqn:nt}
\end{align}
where $|\vec{n};t\rangle$ in the position representation is 
$$|\vec{n};t\rangle = \frac{1}{\sqrt{n_1!\ldots n_L!}} \sum\limits_{\pi \in S_N} w_{s(1)}(x_{\pi(1)},t)\ldots w_{s(N)}(x_{\pi(N)},t).$$
Here $s(n)$ is a~sequence for which exactly $n_l$ terms take a~value of $l.$
This construction defines a~variational manifold embedded in the full Hilbert space of the problem. 
Observe that  all the particles at a~given site  occupy the same time-dependent mode.
Thus by construction they are in a~separable state where multiparticle entanglement is absent.

The state $|\vec{n};t\rangle$ depends on time by the time-dependence of Wannier functions $w_i(\vec x,t).$ Thus creation and annihilation operators for bosons are also time-dependent and are denoted by $\hat{b}_k(t)$ and $\hat{b}^\dagger_k(t)$. At any time $t$ a~commutation relation $[\hat{b}_k(t),\hat{b}^\dagger_q(t)]=\delta_{kq}$ is fulfilled. In the complete analogy to an~ordinary Bose-Hubbard Hamiltonian, one may define, a~time dependent Bose-Hubbard model \cite{Sakmann2011}
\begin{eqnarray}
\hat{H}_\mathrm{V}&=&\sum_{k=1}^L\left[-J_{kk+1}(t)\hat{b}^\dagger_k(t)\hat{b}_{k+1}(t)+h.c.\right.\nonumber \\
&&\left.+E_k\hat{n}_k(t)+\frac{1}{2}U_{kkkk}(t)\hat{n}_k(t)(\hat{n}_k(t)-1)\right],
\end{eqnarray}
where $J_{kk+1}(t)$, $E_k(t)$, $U_{kkkk}(t)$ are a~hopping integral, an on-site energy, and an interaction energy defined respectively as: 
\bea
J_{kk+1}(t)&=&-\int w^*_k(x,t)\hat{h}(x)w_{k+1}(x,t)\mathrm{d}x,\\
E_k(t)&=&\int w^*_k(x,t)\hat{h}(x)w_{k}(x,t)\mathrm{d}x, \\
U_{kkkk}(t)&=&g\int w^*_k(x,t)w^*_k(x,t)w_k(x,t)w_{k}(x,t)\mathrm{d}x.
\eea
A~standard formulation of the time-dependent variational principle (TDV) assumes a~minimization of  the action functional (Lagrange multipliers $\mu_i$ are added to preserve the orthonormality of the variational Wannier functions):

\begin{eqnarray}
S({C_{\vec{n}}},{d_k^\alpha})&=&
\int\langle\psi|\hat{H}_{\mathrm{V}}-i{\partial_t}|\psi\rangle\nonumber\\
&-&\sum_i \mu_i(t)(\langle w_i(x,t)|w_i(x,t)\rangle-1)\ \mathrm{d}t.
\end{eqnarray}

Evolution equations for a~vector 
$|\dot{w}_k(t)\rangle$ and Fock space coefficients $C_{\vec{n}}$, follow:

\begin{align}\label{eqn:evo_sakmann}
i|\dot{w}_k(t)\rangle&=\hat{P}_k(x,t)\left[\sum_{l=k\pm1}^{M}\frac{\rho_{kl}(t)}{\rho_{kk}(t)}\hat{h}(x)|w_l(x,t)\rangle\right. \nonumber \\ 
&\left.+\hat{h}(x)|w_k(x,t)\rangle+\frac{\rho_{kkkk}(t)}{\rho_{kk}(t)}U_{kk}(x,t)|w_k(x,t)\rangle\right],\\
i\dot{C}_{\vec{n}}(t)&=\sum_{\vec{n}'}\langle\vec{n}|\hat{H}_{\mathrm{V}}(t)|\vec{n}'\rangle C_{\vec{n}'}(t),
\end{align}
where $\hat{P}_k(x,t)$ are projection operators: 
\be \hat{P}_k(x,t)=\sum_{\alpha=1}^{{\cal{N}}_\mathrm{V}}|w_k^\alpha(x)\rangle\langle w_k^\alpha(x)|-|w_k(x,t)\rangle\langle w_k(x,t)|\nonumber,\ee with
 \bea \rho_{kl}&=&\langle\psi(x,t)|\hat{b}^\dagger_k(t)\hat{b}_l(t)|\psi(x,t)\rangle\nonumber,\\ \rho_{kkkk}&=&\langle\psi(x,t)|\hat{b}^\dagger_k(t)\hat{b}^\dagger_k(t)\hat{b}_k(t)\hat{b}_k(t)|\psi(x,t)\rangle \nonumber\eea and $U_{kk}(x,t)=g|w_k(x,t)|^2$.

Working out explicitly all the terms of  Eq.~(\ref{eqn:evo_sakmann}) that couple different components of a~vector $d_k(t)$ yields 
\begin{align}
\dot{d}_k^a(t)=(\ldots) + i\sum_{\alpha,\beta,\gamma}^{{\cal{N}}_\mathrm{V}}U^{a\alpha\beta\gamma}_{kkkk}(t)d^{\alpha *}_k(t) d^\beta_k(t) d^\gamma_k(t) \nonumber\\ 
-i\sum_{\alpha\beta\gamma\delta}^{{\cal{N}}_\mathrm{V}} U_{kkkk}^{\alpha\beta\gamma\delta}(t)d^{\alpha *}_k(t)d^{\beta *}_k(t)d^\gamma_k(t) d_k^\delta(t) d_k^a (t).
\label{k1}
\end{align}
The parity symmetry of Wannier functions implies that $U^{\alpha\beta\gamma\delta}_{kkkk}\neq0$ only if sum $\alpha+\beta+\gamma+\delta$ is even. Now if all $d_k^\alpha$ for even(odd) $\alpha$ are set initially to $0$, then $\dot{d}_k^\alpha=0$ for all $t$.

\section{Simulations}

The MBH model as an approximation of the true Hamiltonian (\ref{eq:Hx}) is not very practical. Even restricting the single site space considering states with maximal occupation of a~few bosons per lattice site, the total dimension of that space grows exponentially with the number of Bloch bands, ${\cal N}$, used. The TDV  
approach reduces that dimension dramatically, potentially leading to a great improvement of the efficiency.  We shall compare below both approaches on a~simple model system
consisting of 4 lattice sites among which a~total number of 6 bosons has been distributed. We assume periodic boundary conditions. On-site energies, hopping integrals and interaction energies are calculated using Wannier functions for this four site lattice.

Typically we consider first 3-5 bands for the MBH model. For the TDV simulation of this system we consider a~sufficient number of Bloch bands, ${\cal N}_V$, to allow for convergence of the variational Wannier functions, as this increases the total computational cost very little (usually convergence is reached for 3-5 bands).

The energy is measured in the units of recoil energy, $E_\mathrm{r}={h^2}/{2m(2a)^2}$ with $a=\lambda/2$ being the lattice constant. The depth of the lattice is typically set by us to $s=10E_\mathrm{r}$. Simulations of the TDV model are made with Mathematica's NDSolve function.

\subsection{Ground state}
An energy of a~ground state can be used as a~simple quantity enabling one to compare the accuracy of  state representation over various variational manifolds.  It
has been calculated numerically for different  coupling constant $g$ using up to 5 Bloch bands in both approaches.
For the special case of a~single Bloch band, ${\cal N}=1,$ both methods reduce obviously to the same standard BH model and lead to the same ground state energy. It is no longer true when more bands are taken into consideration. Let us denote ${\cal N}_M$ the number of bands used within MBH (keeping ${\cal N}_V$ for the variational approach).

\begin{figure}
\includegraphics[width=8cm]{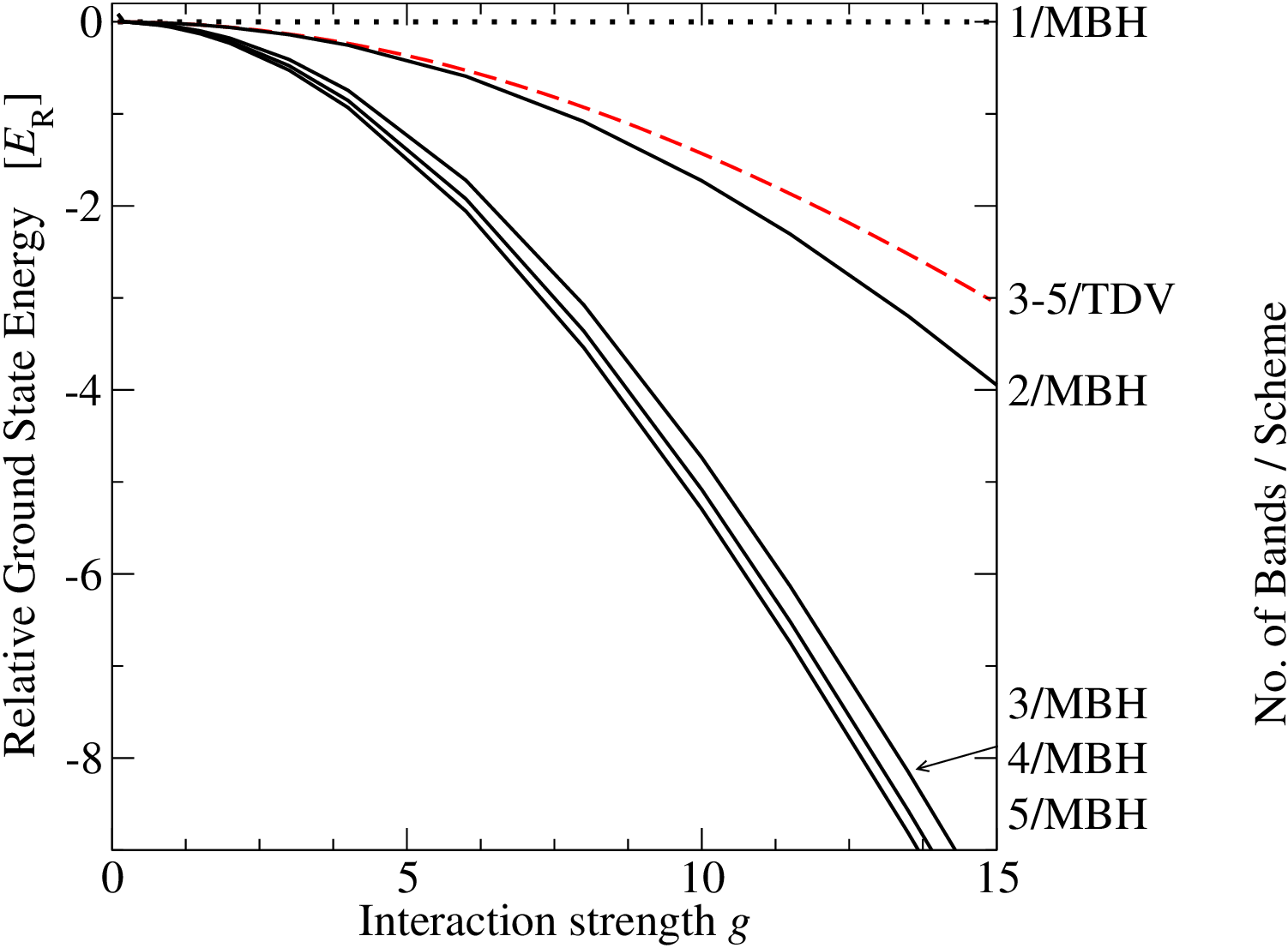}
\caption{(color online) Ground state energy for 6 particles in a~4-site system with periodic boundary conditions calculated within MBH model for ${\cal N}_M=1,2,3,4,5$ Bloch bands included (black curves with number of band indicated). Results for the variational ansatz of \cite{Sakmann2011} are shown as a~dashed red curve - they converge for ${\cal N}_V>2$. The energy is shown with respect to the ground state energy of the standard Bose-Hubbard model as a~function of the coupling constant $g.$  Clearly for $g>1$ the TDV approach based on (\ref{eqn:wannierSak}) fails to approximate the ground state energy.}
\label{fig:gsene}
\end{figure}

In Fig.\ref{fig:gsene} estimates for the ground state energy are presented. Notice that the TDV ansatz leads to the apparent convergence of the estimated ground state energy quickly (for ${\cal N}_V\ge 3$) even for large values of the interaction constant. Observe also that   already ${\cal N}_M=2 $ is sufficient  in MBH approach
to yield lower estimate for the energy. Here for large  $g$ a~slow convergence with increasing ${\cal N}_M$ is observed. On the other hand for small $g<1$ TDV as well as MBH predictions become close to the standard BH model pointing out its region of validity.

The failure of TDV for larger $g$ indicates that even the ground state in the model involves significant entanglement between particles, the feature absent in the variational ansatz (\ref{eqn:wannierSak}) where all the particles at a~given site are in the same, variationaly chosen Wannier state.

\subsection{Time evolution}
Let us now compare time evolution in both approaches. Rather than starting this evolution from the appropriate ground states (which may differ significantly - see above) we   consider model initial states that enlight the differences between MBH and TDV results.
The time evolution in the TDV model is performed by solving numerically the system of differential equations (\ref{eqn:evo_sakmann}). For the MBH  a~many body Schr\"odinger equation is solved (which is easy for our small model system).

We study  evolution of the system using both approaches in three cases: with constant interaction strength but inhomogeneous distribution of bosons over sites, with linearly quenched coupling constant, and with oscillating one. Time-dependent $g$ may be realized by varying the  magnetic field $B(t)$ close to Feshbach resonance. The alternative would be to vary the lattice depth $s$. That, for rapid changes of $s(t)$ may lead to additional effects \cite{Lacki2013} which we want to avoid presently for clarity.

\subsubsection{The inhomogeneous distribution of particles}
We performed numerically the evolution of the system with initial state being a~Fock state, containing the initial distribution of 6 particles over 4 lattice sites as: $(2,2,1,1).$  Particles in sites 1,3,4 are confined initially to the lowest Bloch band, while two particles localized in the site 2, are either also put in the lowest Bloch band or first or second excited band. During the numerical integration of the time-dependent Schr\"odingier equation populations of all four lattice sites are monitored.
\begin{figure}
\includegraphics[width=8cm]{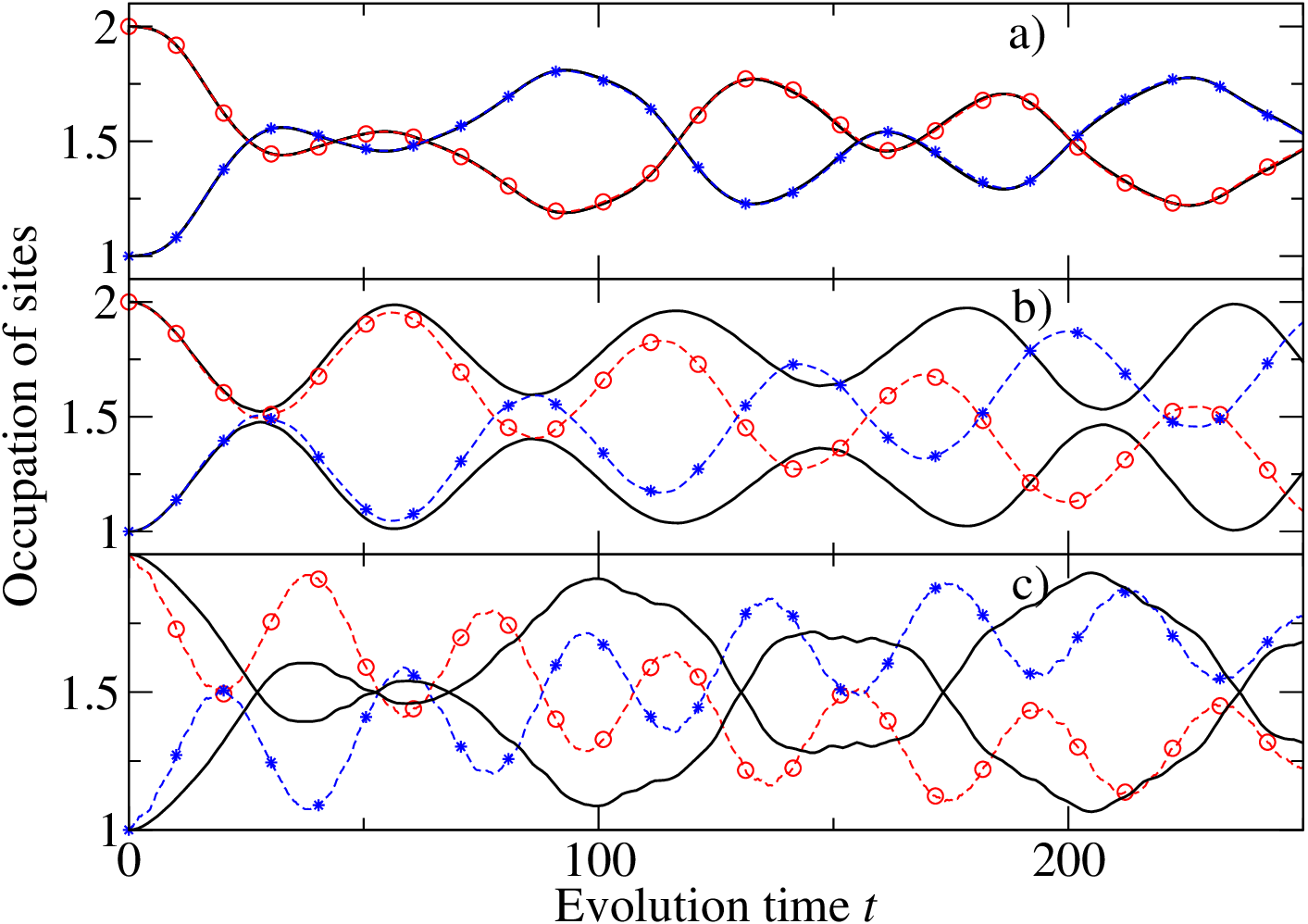}
\caption{(color online) Population of lattice sites in time for the initial Fock state $|2,2,1,1\rangle$ (all particles in the lowest band) for different interaction strengths: panel a) $g=0.2$, panel b) $g=2$, panel c) $g=4$. Results from TDV approach are represented by black solid curves, MBH predictions are shown as colour curves with stars and circles. First two and second two sites are equivalent due to periodic boundary conditions assumed.}
\label{fig:2211wan0}
\end{figure}

\begin{figure}
\includegraphics[width=8cm]{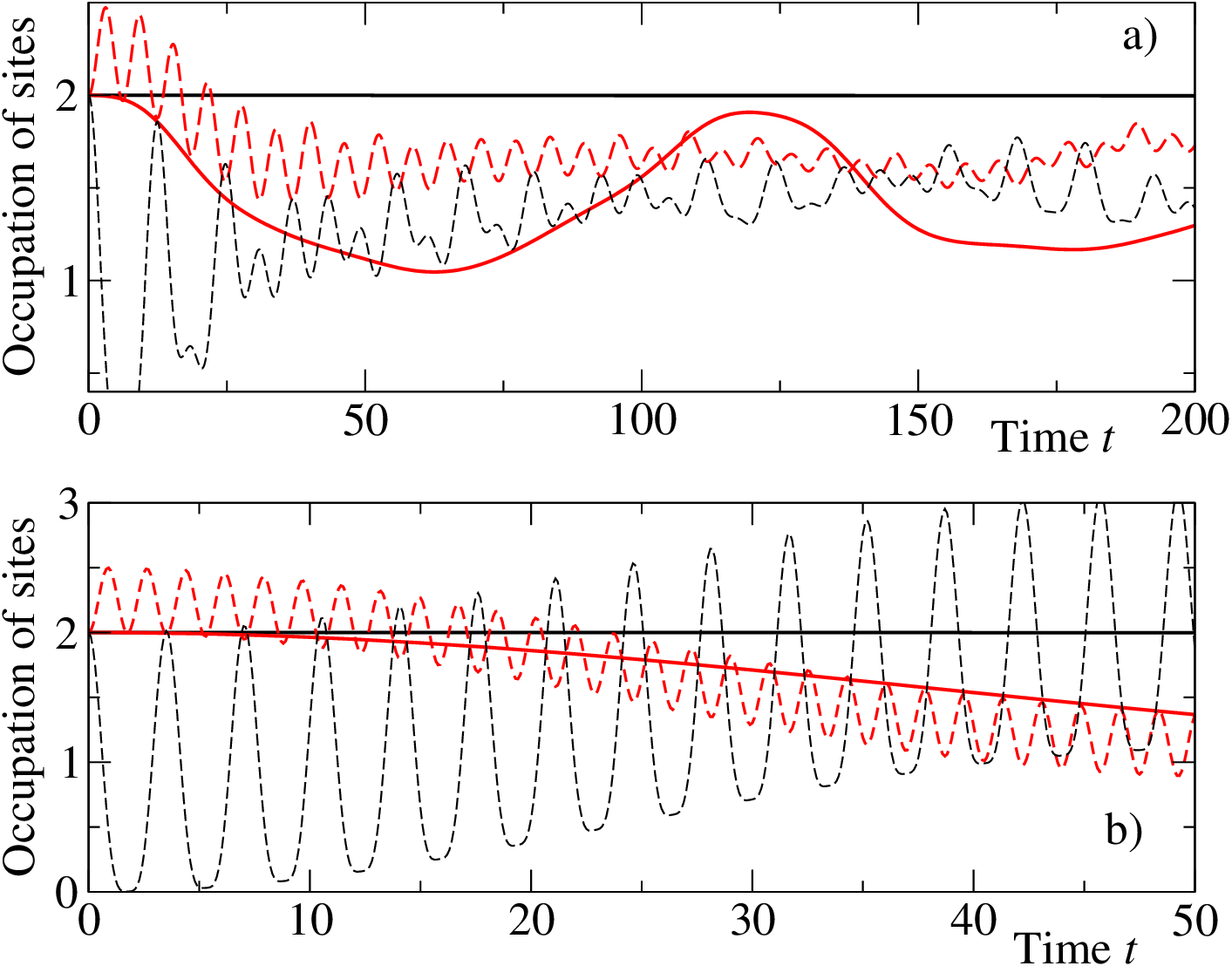}
\caption{(color online)  Populations of different lattice sites in time for the initial Fock state $|2^{\small+},2,1,1\rangle$ (panel a), and $|2^{\small+\!+},2,1,1\rangle$ (panel b), where "+" ("++"), denotes occupation by both particles of the first (second) excited Bloch band.  Black curves show occupation of site 1, red curves of site 2. Solid (dashed) lines represent results obtained wit the help of TDV (MBH) models. The site in which particles originally resided in the excited band is not depleted at all within the TDV approach if the excited band has opposite parity (case shown in top plot, with $g=0.2$) or if the atoms are non-interacting (bottom, $g=0$). 
}
\label{fig:2211wan1}
\end{figure}

Let us consider first the case when all the particles were put in the lowest Bloch band. If the interaction strength coupling constant $g$ is small enough ($g\approx0.2$) results obtained using both methods are virtually the same (compare Fig.~\ref{fig:2211wan0}). Due to the symmetry of the system two sites having initially single occupancy are equivalent (the same holds for initially doubly occupied sites). Thus only two distinct curves appear in the plot with population between sites being transferred in an oscillatory manner. For larger $g$ the
predictions of both approaches  start to diverge for longer times but for short enough times remain similar and TDV can be used in this regime to get approximate results (eg. for $g=1$ for the time of one oscillation). But when $g$ is  large, results for both methods differ considerably in time shorter than a~single oscillation (as for $g=4$). All the results presented are obtained using 3 Bloch bands in MBH. For TDV method we use up to 5 bands (we checked that the results are converged with respect to number of bands in both approaches).

If the two particles are put in the first excited state in the site 2 initially the differences become much more striking. The variational approach is incapable to show any transport of particles that occupied the first excited band into the adjacent sites. This is obviously incorrect and results from the restriction of TDV ansatz in which all particles at a~given site occupy the same time dependent Wannier orbital.. The MBH approach has no such a~restriction.

In the case of particles put in the second excited state when interactions are set to zero, tunnelling in TDV model also does not appear. Only in the presence of interactions some transport between sites  is restored but obviously it has a~different, interaction based origin. In effect  the  simulations in MBH and TDV approach show different results.

The difference between TDV and MBH results can be understood using a~simplified case of two particles in two wells system. Assume that initially the time dependent Wannier function in the first site is purely a~ground state ($w_1=w_1^1$), while in the second site in an excited state ($w_2=w_2^\epsilon$). The tunnelings between such Wannier states vanish $J_{12}=J_{21}=0$. The transport between sites may result from interactions only provided the bands are of the same symmetry.  For opposite symmetry of bands the parity rule discussed in the context of Eq.~(\ref{k1}) implies vanishing coupling between sites. Then  $i\dot{C}_{\vec{n}}(t)=\langle\vec{n}|H_{\mathrm{V}}(t)|\vec{n}\rangle C_{\vec{n}}(t)$ gives only a~phase change and occupations remain constant.

\begin{figure}
\includegraphics[width=8cm]{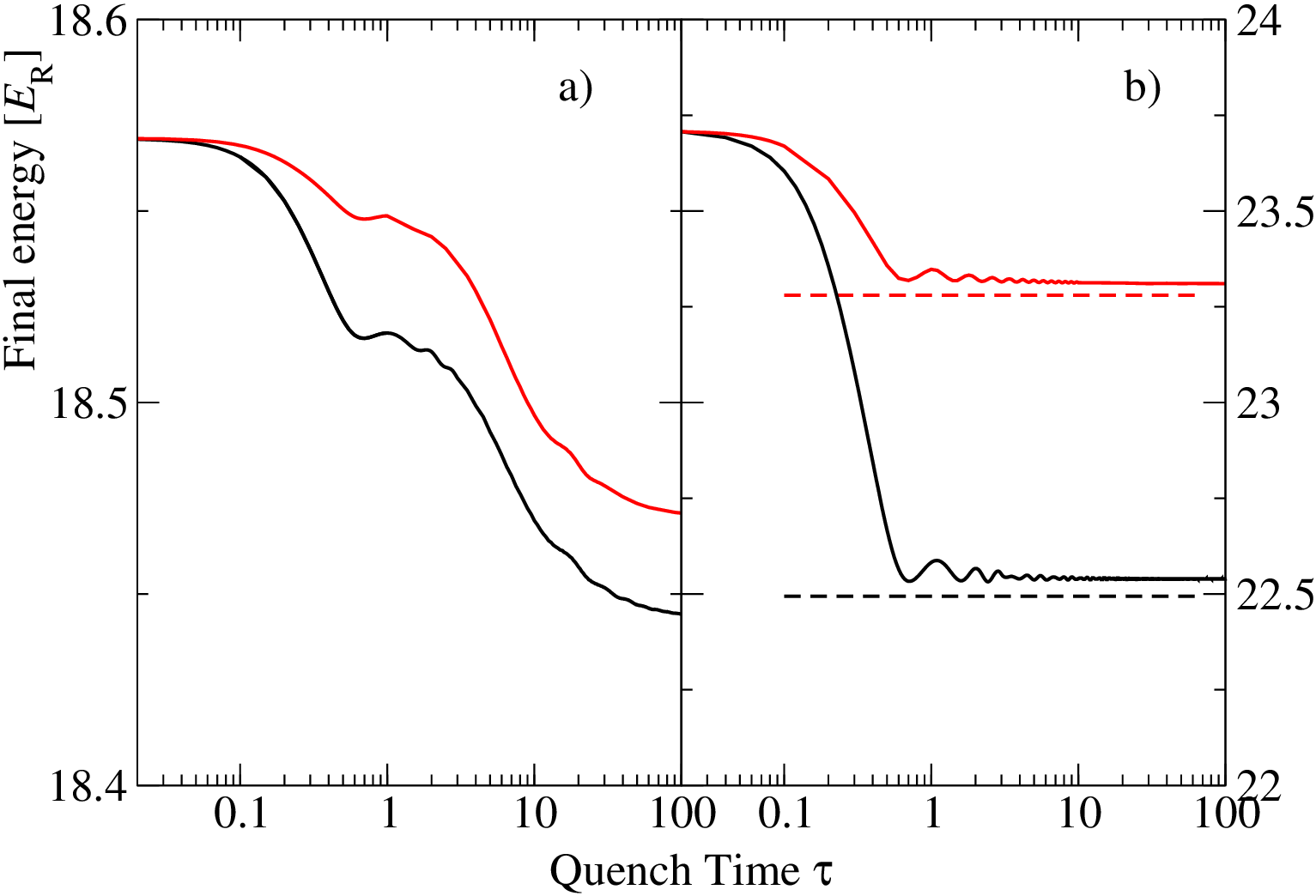}
\caption{(color online)  Final energy after a~linear quench of interaction strength $g$ during time $\tau$, in panel a) starting from $g_{ini}=0.2$ up to $g_{fin}=1.0,$ in panel b) $g_{ini}=1.0,$  $g_{fin}=5.0$ Red (upper) solid curves show results obtained using the TDV method (for 5 bands), while black thicker lines correspond to the simulation using MBH model (with 3 bands). Horizontal dashed lines show ground state energies for corresponding methods (see text).
\label{fig:quench}
}

\end{figure}

\subsubsection{Quench}
Consider a~simple quench scenario, a~linear change of strength of two particle interactions from initial value $g_{ini}$ to $g_{fin}$ over time $\tau.$ Initial state has been prepared in the ground state of the single band BH model with $g=g_{ini}$. This assures the same initial state for both methods. Numerical solution of the time-dependent Schr\"odingier equation is performed by means of Runge-Kutta numerical scheme both for the MBH model and the TDV approach.  Fig.\ref{fig:quench} illustrates two cases: $(g_{ini},g_{fin})=(0.2,1)$ and $(g_{ini},g_{fin})=(1,5).$  For a~sufficiently slow  quench, the final energy of the system after the quench is close to the ground state energy of the Hamiltonian with $g=g_{fin}.$ Note that we have started from a~good approximation of the ground state for small initial $g=g_{ini}$, but not exactly from a~ground state, so we do not expect to reach the ground state at  the end of the quench even in  $\tau\to\infty$ limit. Obviously, however, these final ground state energies give the lower bound for the  energies possible to obtain using both methods. It is clear from Fig.~\ref{fig:quench} that indeed the difference between predictions for the final energy  is largely due to the inability of the TDV ansatz to reproduce the ground state energy accurately for large values of $g.$
 
One may observe, however, that the excess energy over the corresponding ground state as well as the shape of energy versus quench time dependence is quite similar in both MBH and TDV approaches. 

\subsubsection{Modulation}

Periodic modulations of system parameters (e.g. optical lattice depth or the interaction strength)  serves as a~mean to transfer the energy to cold atomic system. Sensitivity of the process with respect to the modulation frequency allows to find excitation spectra providing, e.g.,  information about the energy gap in the system 
\cite{Stoferle2004,Iucci2006}  or enabling to study the multi band interaction effects \cite{Mark2011}. Larger modulation frequencies help to control effective tunnelings \cite{Eckardt2005},
resonant driving may lead to a~direct population of excited bands \cite{Sowinski2012,Lacki2013}. 
Analysis of periodic modulations has been also a~useful theoretical-numerical tool \cite{Zakrzewski2009}, to access to the exited states  of BH-like systems. 

Here we consider a~periodic modulation of the system by varying  the interaction coupling constant: $g(t)=g_{0}+g_{mod}\sin\omega t.$ Specifically we take $g_0=1, g_{mod}=0.1.$  The depth of the lattice potential is assumed to be $s=25E_R,$ deep in the Mott regime with vanishing tunnelling. Then the analysis may be reduced to a~single site, in which we put 2 particles. The initial state is a~single Bloch band ground state. This initial condition has an overlap over 98\%  on the energy minimum state in the variational manifold  and a~similar value on  the MBH ground state. At characteristic, resonant frequencies one expects that strong Rabi oscillations occur manifesting efficient excitation of excited bands. To detect the resonance, it is sufficient to measure the depletion of the initial state. In parallel to \cite{Sowinski2012} we define a~transfer efficiency function:

\be
D(\omega) = 1-\inf\limits_{t\in[0,T]} |\langle \psi(0) | \psi(t) \rangle |, 
\ee
where $T$ is a~fixed (long) evolution time.  

\begin{figure}
\includegraphics[width=8cm]{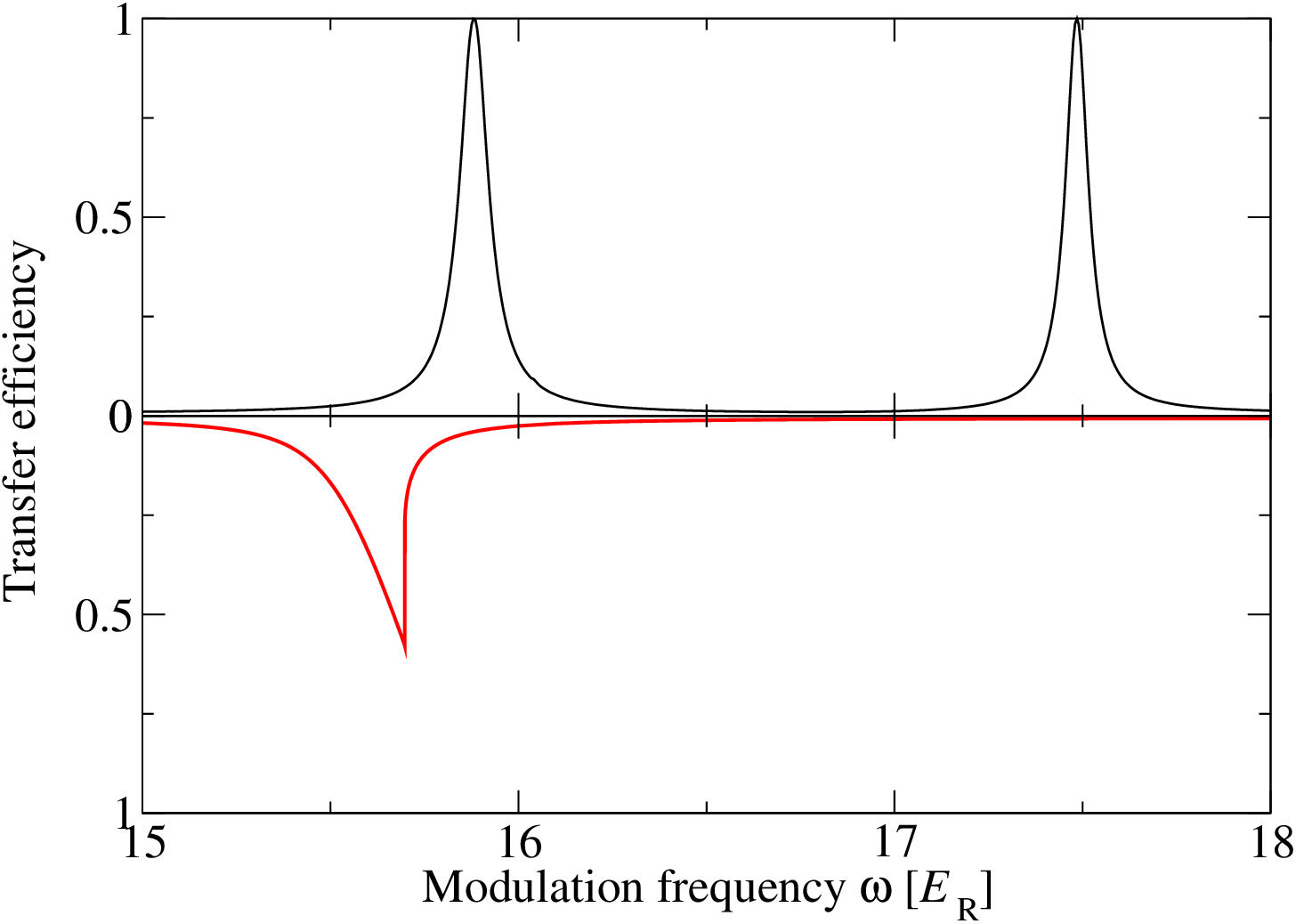}
\caption{(color online) Transfer efficiency from the ground state during a~modulation of total duration $T=400 \hbar/{E_R}$. Top panel shows MBH results, lower panel (in the mirror image) corresponds to TDV model calculations.}
\label{fig:modul}
\end{figure}
 
The depletion as a~function of the frequency of modulation is shown in  Fig. \ref{fig:modul}. The MBH shows two prominent peaks at $\omega\approx 15.9E_{\mathrm R}$ and $\omega\approx 17.5E_{\mathrm R}$. 
The latter may be identified as a~double occupancy of first excited Bloch band. This is strictly forbidden in TDV model: as mentioned before, occupation of Bloch bands $2,4,6,\ldots,$ when starting from initial state containing particles populating  $1,3,5,\ldots$ bands (in our case only the first band), is not possible. Thus the corresponding peak in TDV approach is missing.

Another noteworthy feature of Fig. \ref{fig:modul} is a~noticeable, though small, shift of the single (in this frequency range) absorption peak in the TDV case. This peak is identified in the MBH model as the  interaction-induced promotion  of two particles to the second-excited Bloch band. The TDV dynamics shows a~similar behaviour with significant population of the  second excited band.  The striking asymmetry of the TDV peak (compare Fig.~\ref{fig:modul}) with sharp drop (be aware of the mirror image) in the right hand side is an unexplained peculiarity of TDV approach numerics. This is not a numerical unstability as checked by high precision arithmetics using Mathematica code. 

The state in which two particles occupy the same site: one in the lowest Bloch band, the other in the second-excited band is: $\psi_{13}=\hat{b}_k^1{}^\dagger \hat{b}_k^3{}^\dagger |\Omega\rangle$ is not representable by the variational ansatz. Indeed, such a~state is a~maximally entangled state of two particles. The maximal overlap of $\psi_{13}$ over a~variational product state: $|\langle \psi_{13} | \frac{1}{\sqrt{2}}(\alpha b_k^1{}^\dagger+\beta b_k^3{}^\dagger)^2|\Omega\rangle|, |\alpha|^2+|\beta|^2=1$ is $\frac{1}{\sqrt{2}}$ and is reached when $\alpha=\beta=\frac{1}{\sqrt{2}}.$ Such a~state has an energy of $E_1+E_3$ just as the state $\psi_{13}.$ This is quite accurately represented in the simulations: the position of the MBH peak is 15.9 $E_r$ while the TDV model leads to a~highly asymmetric  peak situated at 15.7 $E_r.$ Presumably this shape reflects the drawback of the oversimplified variational space used by the ansatz.(\ref{eqn:wannierSak}). 

\section{Generalization of TDV method}

The variational approach fails in the situations described in this paper largely due to a~large truncation of the Hilbert space, a~truncation denying any possibility for the on-site entanglement to be present in the system. This may be  to some extent improved by introducing more variational bands in the TDV model, leading, however, to a~further complication of the model. Hopefully, in some cases, the number of bands may be kept rather small, allowing for a~reasonable computational efficiency.
For example, for modulation spectroscopy, allowing for just one additional variational band would include the state $a_1^\dagger a_3^\dagger |\Omega\rangle,$ coupled by a~resonance  to the ground state, in the variational space. Excitations of these type dominate modulation spectra \cite{Lacki2012a,Lacki2013a,Huo2011}.

Let us describe the proposed extension of the TDV method in some detail. In the complete analogy to the single variational band approach we suggest to define $D>1$ variational bands (here we consider $D=2$). The equation (\ref{eqn:wannierSak}) is generalised to

\begin{align}
w_k^{\kappa}(x,t)=\sum^{{\cal N}_{V}}_{\alpha=1}d^{\alpha,\kappa}_k(t)w^{\alpha,\kappa}_k(x),
\text{ for }\kappa=1,\ldots,D
\label{eqn:wannierSak2}
\end{align}
The orthonormality is imposed: $\langle w^\kappa_k(x,t),w^{\kappa'}_k(x,t)\rangle=\delta_{\kappa,\kappa'}.$ To obtain equations for the time evolution, time dependent variational principle could be used again. 

\begin{figure}
\includegraphics[width=8cm]{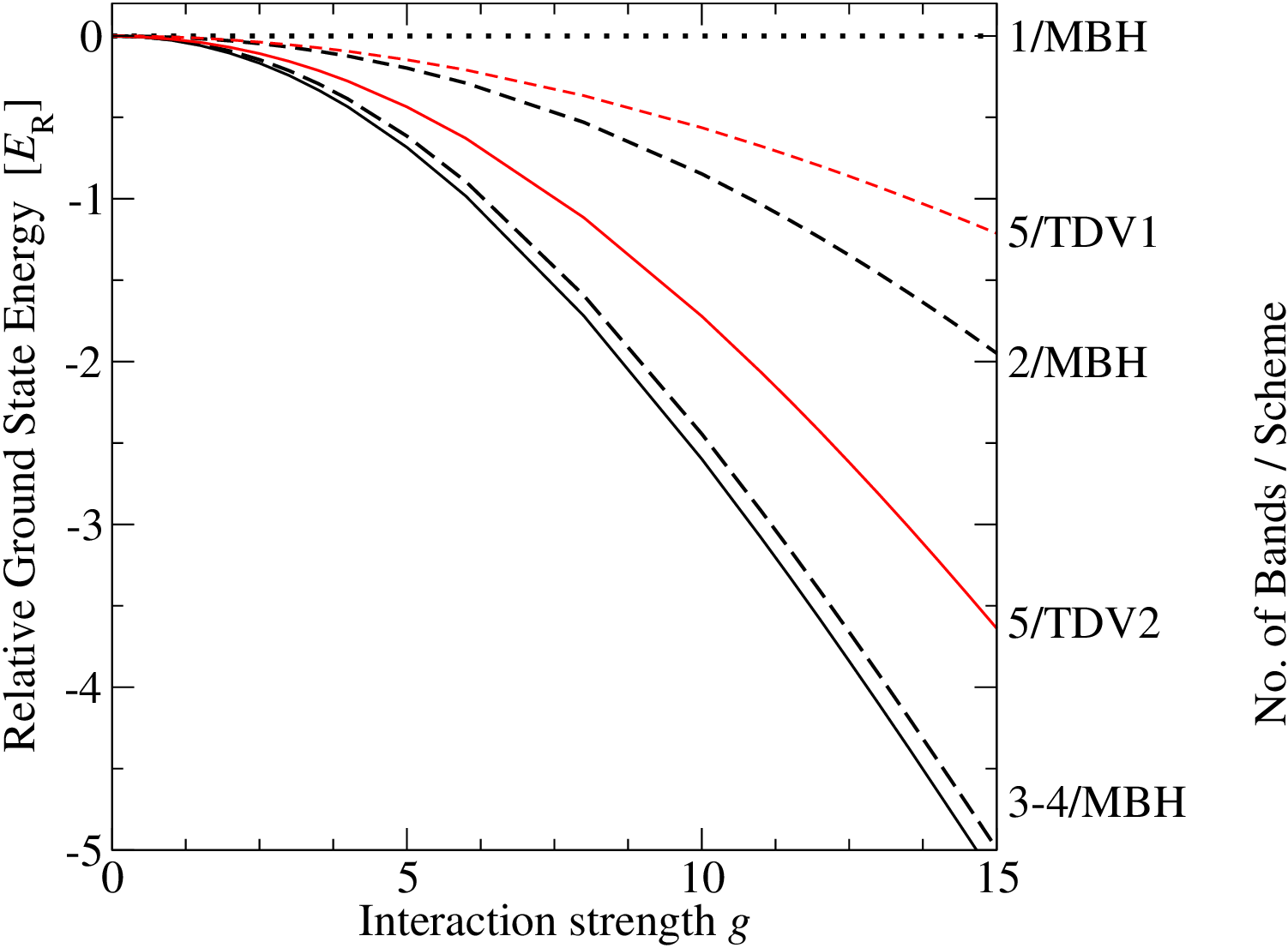}
\caption{(color online)  Ground state energy for 4 particles distributed on 3 sites calculated within MBH model for ${\cal N}_{\mathrm{M}}=1,..,4$ Bloch bands included (black curves). Result for the variational ansatz with $D=1,2$ and ${\cal N}_{\mathrm{V}}=5$ are shown in red. The energy is presented with respect to the ground state energy  of the standard BH model.}
\label{fig:2bandGS}
\end{figure}

Here we test the effect of including $D$ variational bands instead of just one by comparing the ground state energy computation. The energy functional being minimised  reads:
\begin{align}
&\hat{H}=\nonumber\\ &\sum_{k=1}^L\left(\sum_{\kappa,\lambda,\mu,\nu=1}^D\frac{1}{2}U_k^{(\kappa,\lambda,\mu,\nu)}(t)b_k^{(\kappa)\dagger}(t)b_k^{(\lambda)\dagger}(t)b_k^{(\mu)}(t)b_k^{(\nu)}(t)\right.\nonumber\\ & \left.+ \sum_{\mu,\nu=1}^D \left(E_k^{(\mu,\nu)}(t)b^{(\mu)\dagger}_k(t)b^{(\nu)}_k(t)-J_{k,k+1}^{(\mu,\nu)}(t)b^{\dagger(\mu)}_k(t)b^{(\nu)}_{k+1}(t)\right.\right.\nonumber\\ &\left.\left.+c.c.\right) \right)
\end{align}
where 
\bea E_k^{(\mu,\nu)}(t)&=&\int w_k^{\mu*}(x,t)\hat{h}(t)w_k^{\nu}(x,t)\mathrm{d}x\nonumber\\ J_{k,k+1}^{(\mu,\nu)}(t)&=&\int w_k^{\mu*}\hat{h}(t)w_{k+1}^\nu(x,t)\mathrm{d}x\\  U_{kkkk}^{(\kappa,\lambda,\mu,\nu)}(t)&=&\int w_k^{\kappa*}(x,t)w_k^{\lambda*}(x,t)w_k^\mu (x,t)w_k^\nu(x,t)\mathrm{d}x\nonumber
\eea
 we cannot omit one particle cross terms (for example $E_k^{(1,2)}$) because generalised Wannier functions for different variational bands are not formed by eigenstates confined to a~single Bloch band. Such a~TDV model with $D=2$  is compared with MBH model in Fig.~\ref{fig:2bandGS}. For $D<{\cal N}_{\mathrm{V}}<{\cal N}_{\mathrm{M}}$ the TDV space is smaller than the Hilbert space of the MBH model.  If, however  $D<{\cal N}_{\mathrm{M}}<{\cal N}_{\mathrm{V}}$  it is not obvious which approach should be more efficient. 
The complexity of calculations within the limits of the ansatz given by Eq.(\ref{eqn:wannierSak2}) depends largely on $D,$ not on ${\cal N}_{\mathrm{V}},$ thus $D<{\cal N}_{\mathrm{M}}<{\cal N}_{\mathrm{V}}$  situation  is the only one that may result in variational method boosting the efficiency of computation. 
 
Exemplary application of the TDV ansatz for $D=2$ is presented in Fig.~\ref{fig:2bandGS} for 4 particles residing on 3 sites system.
We have found that  $D=2$ leads to a~significant improvement in the estimate for the ground state energy as compared to $D=1$ proposition \cite{Sakmann2011}. In both calculations ${\cal N}_{\mathrm V}=5$. Disappointingly, however, a~comparison with MBH model shows that a~full 3-bands calculation is superior to the TDV ansatz with $D=2$. Thus, while the latter constitutes a~significant improvement over the $D=1$ case,
 it still does not catch the complexity involved in the ground state of the system, in particular for higher interaction strength $g$ values. Seemingly, multiparticle entanglement (missing for $D=2$ that captures two particle entanglement only) becomes important.

It would be desirable to compare $D=1$ and $D=2$ results of TDV approach also for a~slightly larger system of 6 particles on 4 sites as discussed for $D=1$ previously. Unfortunately for $D=2$ the TDV procedure seems to be quite ineffective leading to a~significant slow down of the minimalization procedure due to a~large number of local energy minima in a~highly nonlinear variational equations. This casts a~shadow on a~possible application of TDV approach to really interesting cases.

\section{Conclusions}
We have provided extensive tests of the TDV approach \cite{Sakmann2011} as compared to computationally expensive MBH approach. Unfortunately we have found that TDV approach, despite claims,  provides little alternative for 
moderate and strong interatomic interactions and nontrivial time-dependence of the system. Even extending the TDV approach to a~richer Hilbert space taking into account two particle entangled states helps a~little. That shows that the genuine ground state of strongly interacting bosons in optical lattices constitutes a~clear example of multiparticle entanglement.
Both interaction strength quenches and its modulation may lead to significant excitation of entangled modes - in such cases clearly the TDV approach as advertised by \cite{Sakmann2011} fails to capture the details of the physics involved. Moreover, for periodic modulation of the interaction strength we have observed strange asymmetry in modulation spectra in the TDV approach probably reflecting the fact that the variational space is strongly restricted.
  
\section{acknowledgements}
JZ and JM acknowledge support of the Polish National Science Center grant
DEC-2012/04/A/ST2/00088. MŁ acknowledges support of
the Polish National Science Center by means of project no. 2013/08/T/ST2/00112 for the PhD thesis, and a~research grant  DEC-2011/01/N/ST2/02549 by the same institution. MŁ also acknowledges a~special stipend of Smoluchowski Scientific Consortium ``Matter Energy Future''.
Simulations were carried out at ACK Cyfronet AGH, part of PL-Grid project and on Deszno supercomputer (IF UJ) obtained in the framework of the Polish Innovation Economy Operational Program (POIG.02.01.00-12-023/08).

%

\end{document}